# The Search for Higher $T_c$ in Houston

*(dedicated to Alex Mueller, whose "important breakthrough in the discovery of superconductivity in ceramic materials" in 1986 has changed the world of superconductivity)*

C. W. Chu
Department of Physics and Texas Center for Superconductivity
at the University of Houston

It is a great pleasure to be invited to join the chorus on this auspicious occasion to celebrate Professor K. Alex Mueller's 90$^{th}$ birthday by Professors Annette Bussman-Holder, Hugo Keller, and Antonio Bianconi. As a student in high temperature superconductivity, I am forever grateful to Professor Alex Mueller and Dr. Georg Bednorz "for their important breakthrough in the discovery of superconductivity in the ceramic materials" in 1986 as described in the citation of their 1987 Nobel Prize in Physics. It is this breakthrough discovery that has ushered in the explosion of research activities in high temperature superconductivity (HTS) and has provided immense excitement in HTS science and technology in the ensuing decades till now. Alex has not been resting on his laurels and has continued to search for the origin of the unusual high temperature superconductivity in cuprates.

My first encounter with Alex was through a phone conversation on December 10$^{th}$, 1986, following my short note to him and Georg Bednorz a week earlier (Fig. 1). During the dog days of superconductivity in 1986 when federal funding diminished to a trickle, our work on BPBO did not lead to any $T_c$-jump above 13 K, although it did suggest that oxides with low density of states in the absence of d-electrons could be a possible candidate for higher $T_c$ (as described in my proposal submitted to the National Science Foundation in summer 1986). However, the crucial turning point was the appearance of the paper by J. G. Bednorz and K. A. Mueller (BM)—"Possible high $T_c$ superconductivity in the Ba−La−Cu−O system," *Z. Phys. B*. **64** (1986) [1]. In Fall 1986, I started hopping between Houston to take care of my lab and Washington, D.C., to serve as a program director of the National Science Foundation (NSF), using the federal flex-time scheme and depending on the goodwill of the managements at both the NSF and the University of Houston. I was extremely excited to read the BM paper on November 8 in my lab and felt that we should be able to do better with our experience in oxides. Initially, their paper did not attract the attention it deserved due to the many false alarms of high $T_c$ previously reported, and its modest title did not help either. The first order of business at the time for us was to reproduce their results. We did this very quickly, which made our group Thanksgiving party that year extra special. To share the excitement, I dropped BM a note on December 3, 1986, as shown in Fig. 1. A week later Alex called me from Zurich to thank me. Instead, I thanked him for pointing out a new direction of cuprates for us for higher $T_c$. We briefly touched on the importance of faith in discovery. By this time, we had also had some interesting preliminary results on La-Ba-Cu-O at ambient and high pressure and therefore I mentioned to him that I was full of confidence for a $T_c$ of 77 K, as was also communicated to my friend, Wei-kan Chu, then at University of North Carolina, shown in the Christmas card in Fig. 2. Later, at the 1987 APS March Meeting, Alex confessed to me that he had thought at the time that I was over-optimistic. The rest is the exciting history of high temperature superconductivity.

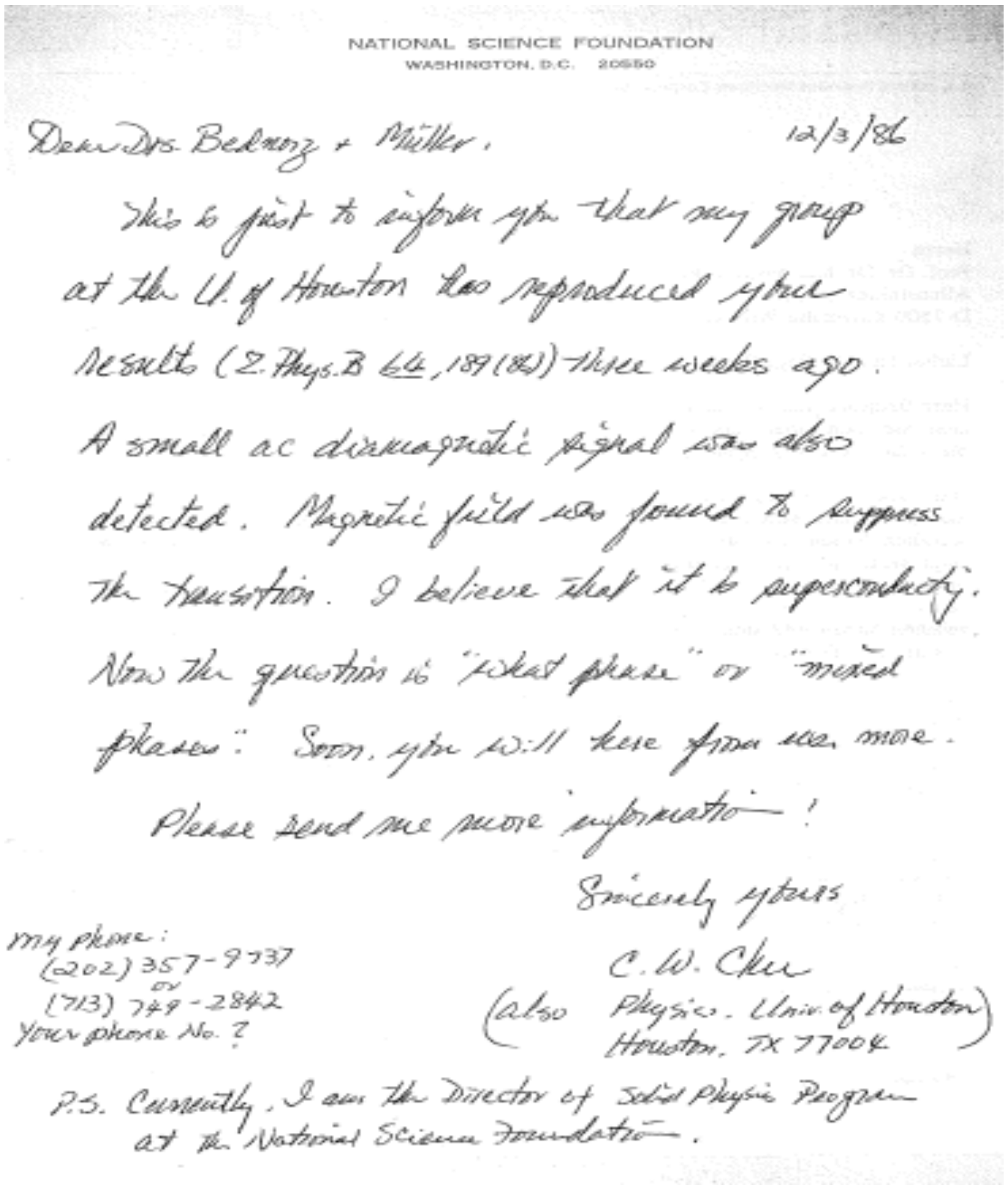
NATIONAL SCIENCE FOUNDATION
WASHINGTON, D.C. 20550

Dear Drs. Bednorz + Müller, 12/3/86

This is just to inform you that my group at the U. of Houston has reproduced your result (Z. Phys. B 64, 189 (86)) three weeks ago. A small ac diamagnetic signal was also detected. Magnetic field was found to suppress the transition. I believe that it is superconductivity. Now the question is "what phase" or "mixed phases". Soon, you will hear from me more. Please send me more information!

Sincerely yours

C. W. Chu

(also Physics. Univ. of Houston
Houston, TX 77004)

my phone:
(202) 357-9737
or
(713) 749-2842
Your phone No.?

P.S. Currently, I am the Director of Solid Physics Program at the National Science Foundation.

**Fig. 1** Provided by Georg Bednorz at the 1997 March APS Meeting in Denver.

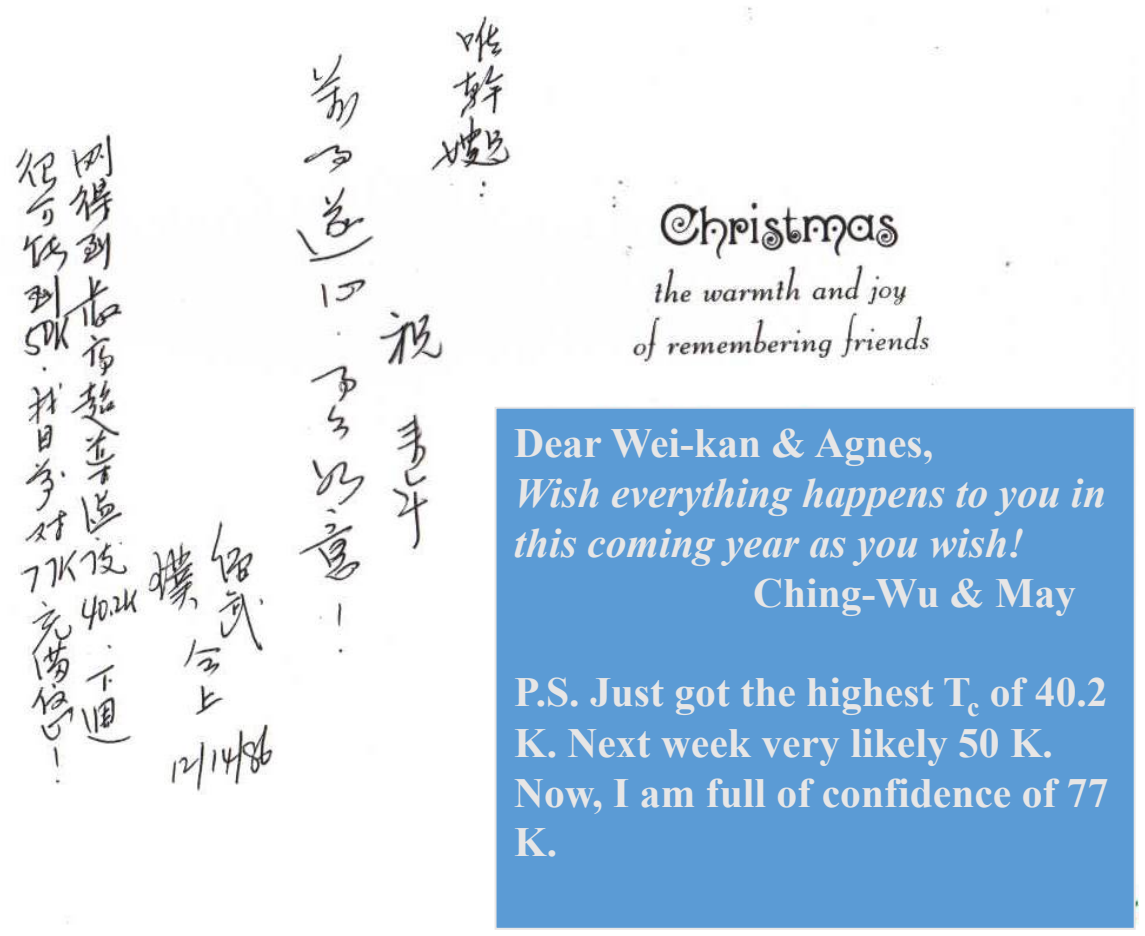

**Fig. 2** A Xmas greeting from C. W. Chu to Wei-kan Chu on December 14, 1986, with translation.

Indeed, in the last three decades, great progress has been made in all areas of HTS research and development: materials, science, and technology. For example, more than 250 stable compounds of cuprates and iron-based pnictides and chalcogenides have been discovered with $T_c$ up to 134 K and 164 K in $HgBa_2Ca_2Cu_3O_7$ at ambient and 32 GPa, respectively (not to mention the most recently reported unstable hydrogen-rich molecular compounds with a $T_c$ up to 203 K under ultrahigh pressures above 200 GPa, to be discussed later); many new unusual phenomena have been observed in these compounds; the electronic, phononic, and magnetic spectra have been determined in many of these compounds; various theoretical models have been proposed to account for the observations; kilometer-long HTS ribbons have been fabricated with respectably high $J_c$; and many prototypes of HTS devices have been successfully constructed and demonstrated with performance better than their non-superconducting counterparts. Unfortunately, to date there exists no theoretical HTS model that is commonly accepted, and commercially viable HTS devices are yet to be made available.

We believe that the most effective way to provide relief to the above impasse is to increase the superconducting material space with higher $T_c$ preferably above room temperature, higher $J_c$ via anisotropy reduction, and a greater number of material and structural types through the discovery of novel compounds. In view of the complexity of the existing HTS materials, a holistic multidisciplinary enlightened empirical approach has been proposed to achieve novel HTS with higher $T_c$ [2].

It is therefore not surprising to find that the search for novel superconducting materials with higher $T_c$ has been an important integral part of superconductivity research ever since its discovery, as shown in Fig. 3. Over the years, the search has lured many great minds and at the same time humbled many. In this congratulatory note, I would like to briefly describe and comment on some of our ongoing efforts at Houston in the search for higher $T_c$. A few examples follow (details of some have been reported elsewhere):

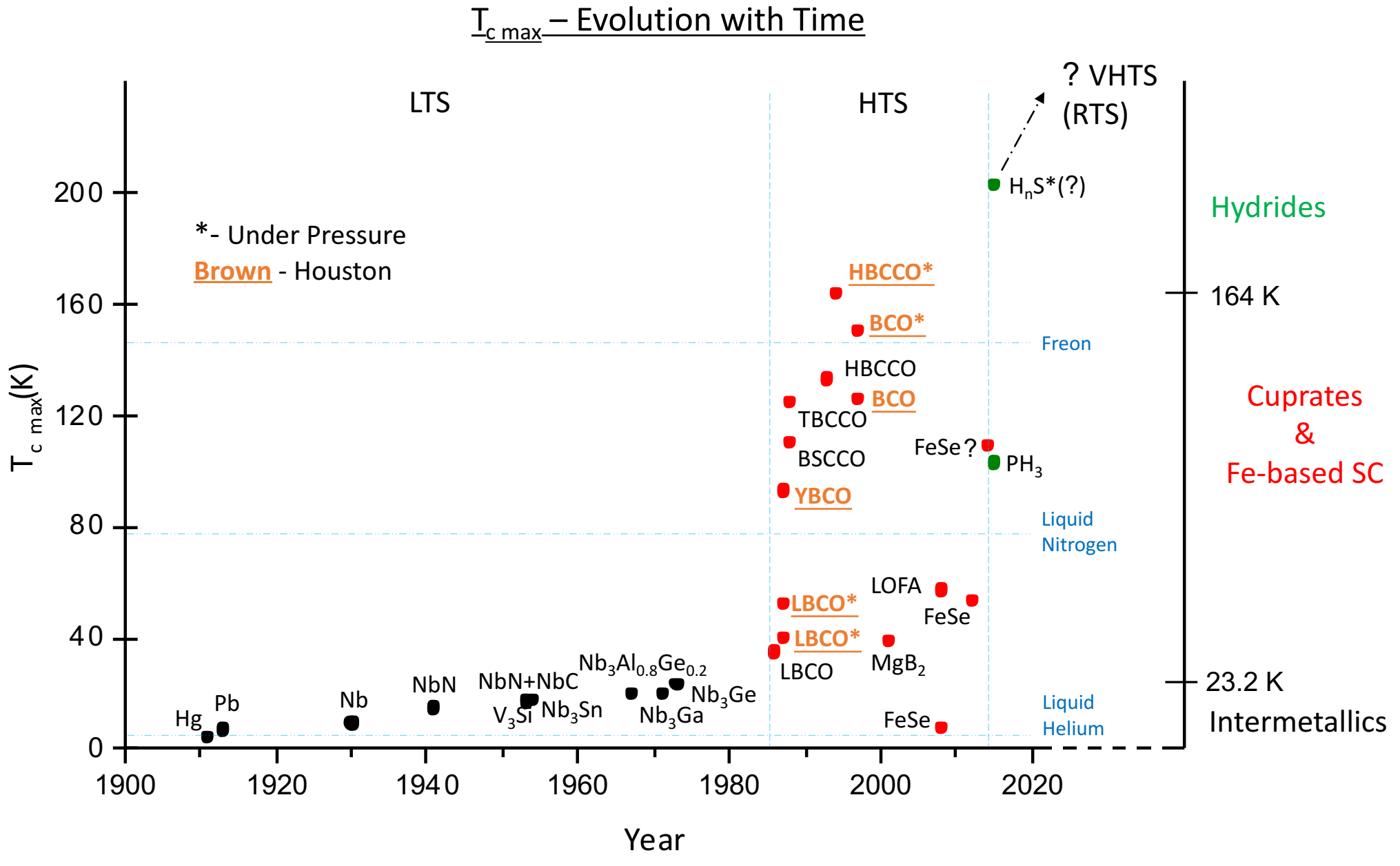

**Fig. 3** $T_{c\ max}$-evolution with time.

## 1. Interface-Induced Superconductivity

Cuprate superconductors with a $T_c$ above 77 K, the liquid nitrogen boiling point, share a common layered structure with multiple block subsystems: an active subsystem, where certain interactions promote charge-carrier pairing, and a charge reservoir subsystem to offer carriers to the former subsystem without introducing defects, similar to that which occurs in a semiconducting superlattice structure. As a consequence, many cuprates, especially those with high $T_c$s, show high anisotropy, or quasi 2D-like features. Excitations with different energy scales are believed to provide the glue for superconducting electron-pairings, although a commonly accepted nature of such excitations for the high $T_c$ is yet to emerge. One way that has long been proposed to achieve enhanced $T_c$s is to take advantage of artificially or naturally assembled interfaces, where soft phonons and/or excitons may occur. In fact, the discovery of the layered cuprate HTS and the great advancement in molecular beam epitaxy growth of perfect thin films have revived the interest in, and accelerated the search for, the interfacial effect in the artificially assembled composite layered compound systems. In addition, naturally assembled metal/semiconductor interfaces have been demonstrated in the BSCCO single crystalline samples by the observation of the Shapiro steps in a microwave environment [3].

Although the idea to achieve high $T_c$ via interfaces is not new, the first serious model analysis was not made until 1973 when Allender, Bray, and Bardeen (ABB) examined the effect of metal/semiconductor interfaces on superconductivity [4]. They found an impressive $T_c$-enhancement effect when the stringent conditions on the interfaces are met. Many experiments were inspired by the prediction. An excellent recent article [5] has summarized numerous experiments that display enhanced superconductivity in artificially formed layered compound systems between two different non-superconducting, one superconducting (sc) and one non-superconducting (nsc), or two different superconducting materials. For instance, superconductivity has been observed in heterostructures of PbTe(nsc)/YbS(nsc) up to 6 K, in Al(sc)/$Al_2O_3$(nsc) up to 6 K vs. 1.2 K in bulk Al, and in $(La,Sr)_2CuO_4$(nsc)/$La_2CuO_4$(nsc) up to 52 K, all suggesting a rather large $T_c$-enhancement effect. Although the exact nature of the enhancement remains unclear, exotic interfacial effects that may facilitate the exchange of excitons have been implied.

The alkaline-earth iron arsenide $CaFe_2As_2$ (Ca122) is the parent compound of a large superconductor family. The superconductivity of many compounds in this family seems to be rather unusual. The most unusual observation is the non-bulk superconductivity up to 49 K induced in Ca122 at ambient pressure by slight Ca-replacement with a rare-earth element, La, Ce, Pr, or Nd [6]. A systematic study on the slightly rare-earth doped Ca122 suggests that the above non-bulk superconductivity detected may be associated with the naturally occurring interfaces associated with defects in the samples [7]. However, direct evidence remains elusive due to complications involved in doping and pressure. As for the superconductivity in undoped Ca122, the situation is even more confusing. Filamentary superconductivity up to ~10 K has been detected sporadically under ambient pressure [8]. Although the related superconducting volume-fraction is much higher only under non-hydrostatic pressure, the origin of the superconductivity remains an open issue. The complications associated with the delicate pressure environment, however, seem to prevent a comprehensive experimental verification.

Later studies show that the complex phase-evolution under pressure can be reproduced through heat treatment [9], which may offer a reproducible, controllable, and reversible environment in which many characterization techniques can be applied to explore the issue, if similar superconductivity can be induced. This motivates our studies. We found [10] that Ca122 quenched from 850 °C or above has a tetragonal structure at room temperature ($P_1$ phase), but transforms to a collapsed-tetragonal phase with a 10% shorter *c* lattice parameter below $T_{cT}$, the T-cT transition temperature [11]. No magnetic ordering is detected over the whole temperature range. The furnace-cooled Ca122 single crystals, on the other hand, exhibit a tetragonal structure at room temperature with a slightly longer *c* ($P_2$ phase). On cooling, they undergo a tetragonal-to-orthorhombic transition (T-O) at $T_N$, closely related to the spin-density-waves transition, to an antiferromagnetic phase. The two transitions, *i.e.* T-cT in $P_1$ and T-O in $P_2$, carry with them distinct resistive and magnetic signatures at $T_{cT}$ and $T_N$, respectively, making their detection rather easy (Fig. 4). We found that neither the $P_1$ nor the $P_2$ phase is superconducting down to 2 K. However, through low temperature annealing at 350 °C for different periods of time t, the $P_1$-phase transforms progressively to the $P_2$-phase as evidenced from the XRD results exhibited in Fig. 5, showing the mixing of two phases in random-stacking without the appearance of a third phase, in agreement with our XRD-simulation. Superconductivity with an almost constant onset $T_c \sim 25$ K appears when the two phases coexist and the superconducting volume fraction scales with the volume of the phase mixture, as displayed in Fig. 6, in agreement with the suggestion of interface-induced superconductivity.

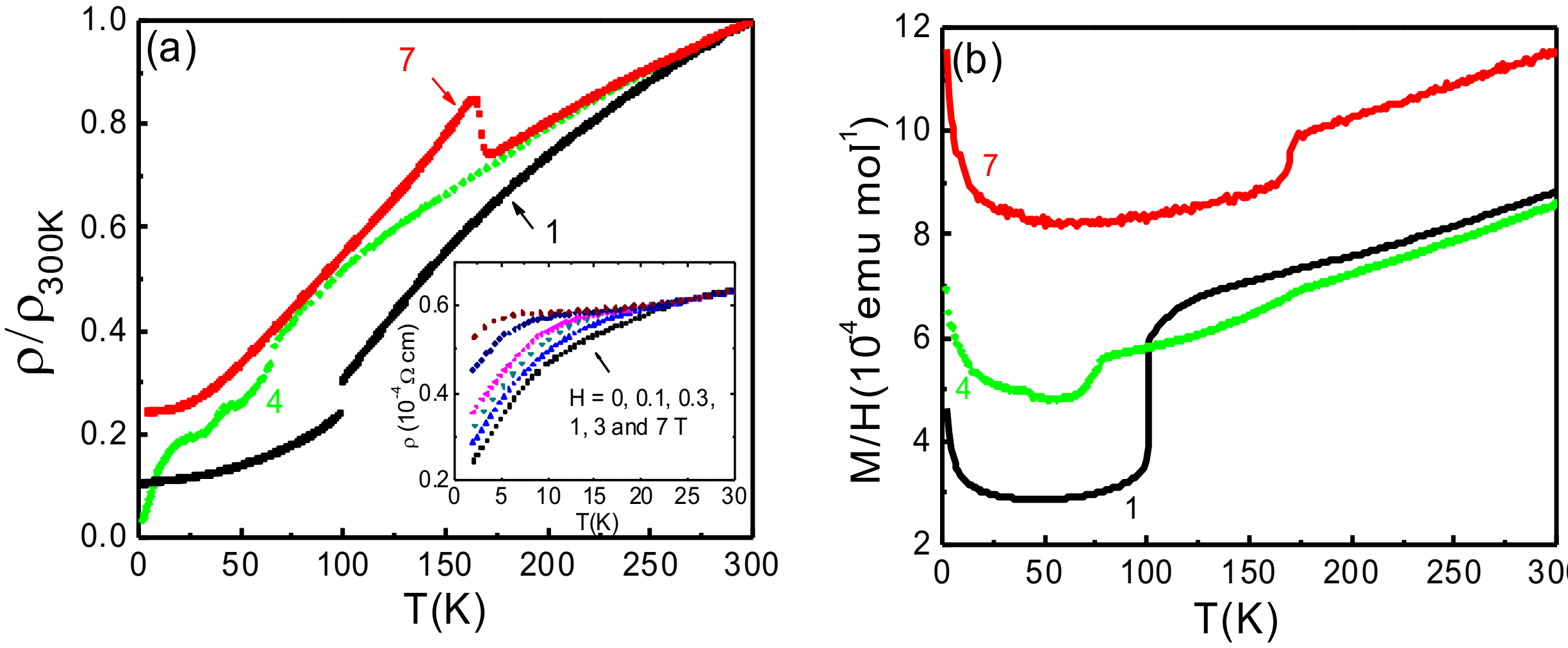

**Fig. 4** #1—$P_1$ phase (quenched from 850 °C); #4—mixed phase (sc, annealed at 350 °C for 12 h); #7—$P_2$ phase (annealed at 350 °C for >16 h). From K. Zhao et al., PNAS 113, 12968 (2016).

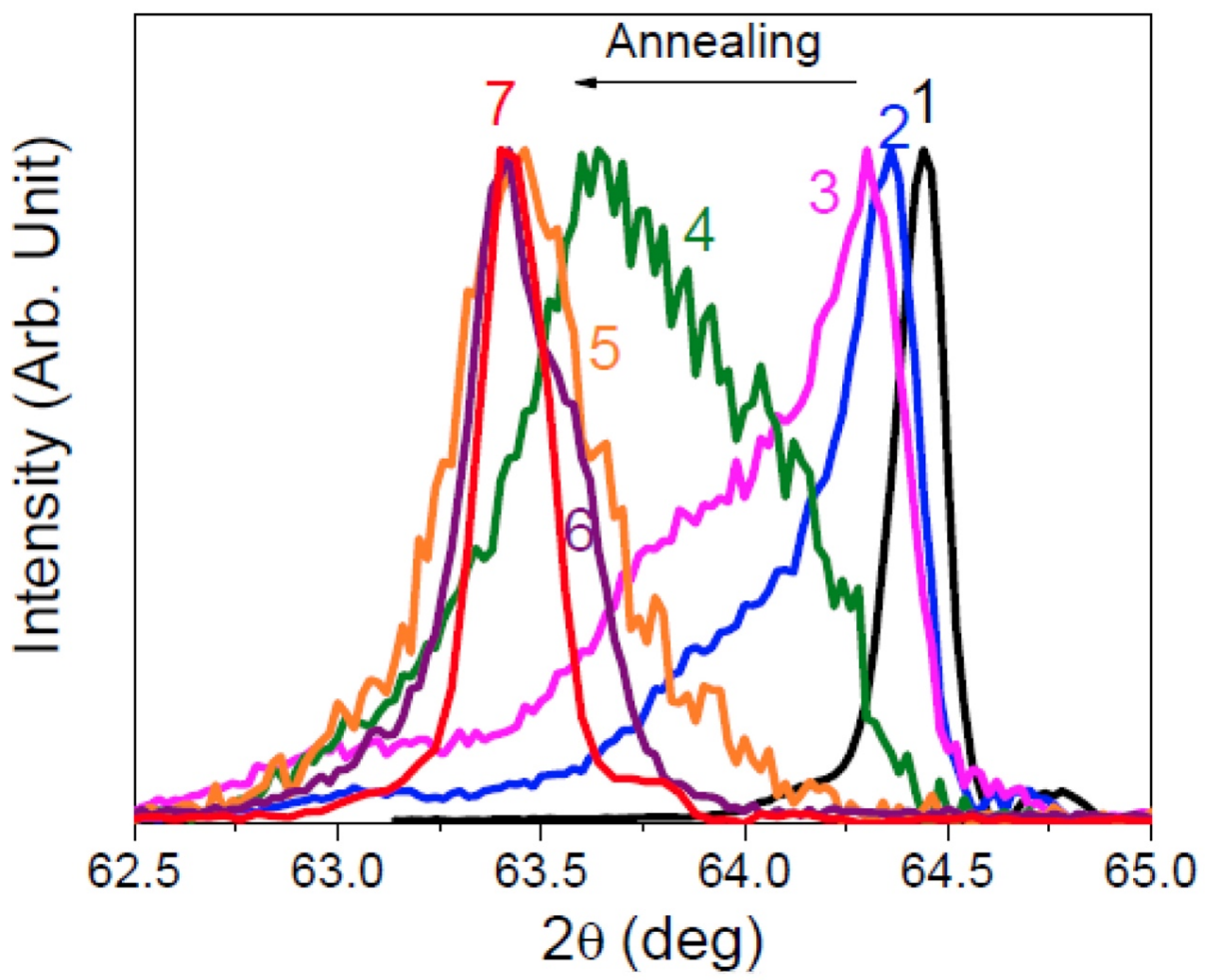

**Fig. 5** XRD for different annealing times at 350 °C. From K. Zhao et al., PNAS 113, 12968 (2016).

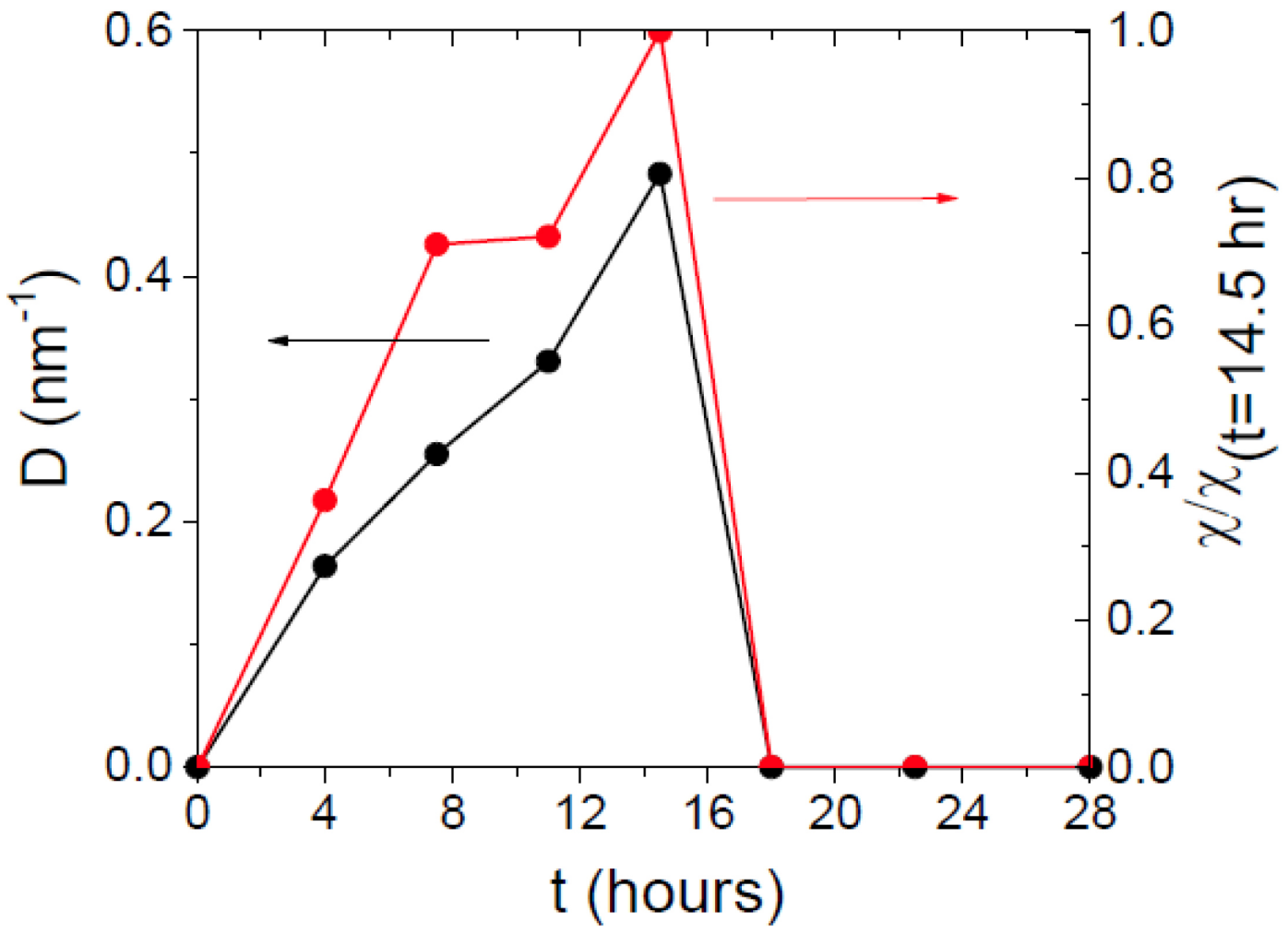

**Fig. 6** Superconducting volume fraction. From K. Zhao et al., PNAS 113, 12968 (2016).

In summary, superconductivity with an almost constant $T_c$ of ~ 25 K can be reversibly induced in the mixed phase region of two non-superconducting phases $P_1$ and $P_2$ in undoped Ca122 by low temperature annealing at 350 °C for different time periods. The XRD data analysis demonstrates that the superconducting samples consist of randomly stacked layers of $P_1$ and $P_2$. The superconductivity volume fraction scales qualitatively with the interface density deduced from the XRD-data. The significant role of interfaces in the 25 K superconductivity is clearly in evidence. Comparison with previous high pressure studies and the small screening volume fraction, however, suggest that a simple strain effect cannot account for the observations. The above observation in undoped Ca122 single crystals represents the most direct evidence for interface-induced superconductivity to date. Similar effect is being explored in other compounds both macroscopically and microscopically. The possible effect of strain during phase mixing will also be further examined.

## 2. Layered Cuprate Superconductors with a $T_c$ above 164 K

Layered cuprate superconductors have dominated the HTS field of study since their discovery. Great efforts have been devoted to the enhancement of their $T_c$ via chemical and/or physical means. Indeed, by controlling these two knobs, $T_c$ has been optimized for each compound or compound family. For example, the peak $T_c$ of the three major cuprate families that superconduct above the liquid nitrogen boiling point, HBCCO [12], BSCCO [13], and YBCO [14], has been observed at 134 K or 164 K, 115 K or 136 K, and 90 K or 92 K at ambient or high pressure, respectively. In spite of the many different models proposed, the general dome-shape of $T_c$-variation with the alteration of the chemical and/or physical parameter appears to be the norm. For example, $T_c$ of a cuprate has been found to follow the universal hole-concentration quadratically [15] and to follow the pressure applied in a similar fashion. In other words, $T_c$ goes up through a maximum and comes back down with chemical dopant or physical pressure applied. Some colleagues even turn philosophical by saying that whatever goes up will come down. This implies that a simple electronic rigid band is in operation within certain ranges of doping and pressure. If this is strictly the case, any attempt to raise the $T_c$ via these means without changing the band structure would appear to be futile. However, the situation can be rather different if a certain kind of electronic transition can be induced by doping and/or pressurizing without triggering the catastrophic collapse of the crystal. Examination of the existing high pressure data on cuprate high temperature superconductors up to 30s GPa by us and others indeed suggests such a possibility to raise the $T_c$ to above the current record of 164 K.

High pressure investigations have been made on the three optimally doped representative members of the families, but only up to 30s GPa, i.e. $Bi_2Sr_2Ca_2Cu_3O_{10}$ (Bi2223) [13], $HgBa_2Ca_2Cu_3O_9$ (Hg1223) [12], and $YBa_2Cu_3O_7$ (Y123) [14]. Under pressures, their $T_c$s all initially increase although at different rates. However, they appear to behave differently at higher pressures. A careful examination shows that certain similarities exist, pointing to the possibility of higher $T_c$. For instance, the $T_c$ of Bi2223, which possesses a unit cell with three $CuO_2$ layers, shows a two-step increase with pressure as shown in Fig. 7. It increases at a moderate rate from 115 K to ~ 128 K at ~ 14 GPa and then decreases to ~ 115 K at 24 GPa but finally increases again up to 136 K at a high rate ~ 1 K/GPa with **no sign of saturation** up to 36 GPa, the highest pressure applied. The second rise in $T_c$ has been attributed to a possible pressure-induced electronic-structure change in the $CuO_2$ inner layer, leading to a possible redistribution of charge carriers and a more favorable competition between pairing and phase ordering in different $CuO_2$ layers in the unit cell. A higher $T_c$ thus appears not just possible but also probable in Bi2223 under higher pressures. Magnetic measurements will tell us if a pressure-induced electronic transition takes place and turns the inner-$CuO_2$-layer antiferromagnetic to activate a possible antiferromagnetic/metallic interfacial interaction predicted to favor a $T_c$-enhancement [16]. The $T_c$-P relation for Hg1223, which has a three $CuO_2$-layer/cell structure similar to Bi2223 has been extensively studied [12] and is displayed in Fig. 8; its $T_c$ increases with pressure, at a rate higher than those of either Bi2223 or Y123, from 134 K to 150 K at ~ 8 GPa, then peaks at 164 K at ~ 32 GPa, and finally decreases only slightly to 160 K at 44 GPa, the highest pressure applied. An increased pressure may allow us to observe the second rise of $T_c$ with pressure in Hg1223 by inducing a possible electronic transition in the $CuO_2$ inner layer, similar to that in Bi2223. In addition, the other two members of the HBCCO family, Hg1201 and -1212, display similar $T_c$-P behavior also shown in Fig. 8, another demonstration of the important similar role of $CuO_2$ layers in the high $T_c$ of the HBCCO family. The ultrahigh pressure study on all three members of the HBCCO family will provide us an opportunity to test the mechanism proposed for the second $T_c$-rise in Bi2223 [13]. In other words, one would not expect to see the second $T_c$-rise in Hg1201 or -1212, which lack the so-called "inner $CuO_2$", at higher pressures, unless different pressure-induced electronic transitions are induced, as discussed below. The $T_c$ of $YBa_2Cu_3O_7$ (Y123) [14] increases initially with pressure from 91 K to ~ 92 K at ~ 8 GPa, then deceases to ~ 60 K at ~ 23 GPa, but drops precipitously to below 2 K beyond 23 GPa, as shown in Fig. 9. This may be attributed to its absence of the inner $CuO_2$ layer in comparison with Bi2223 and Hg1223. However, the re-emergence of another $T_c$ rise in Y123 at higher pressure is not unlikely in view of the recent report of the re-emergence of superconductivity with a higher $T_c$ in Fe-based chalcogenides at higher pressures after the complete suppression of superconductivity by pressure at ~ 10 GPa, as displayed in Fig. 10 [17]. We therefore are investigating the magnetic, resistive, and optical states of Bi2223, Hg1223, and Y123 up to ~ 200 GPa to achieve a $T_c$ above 164 K and examining the possible electronic transition associated with the three $CuO_2$ layers induced by pressure in Bi2223 and Hg1223 and the two $CuO_2$ layers in YBCO.

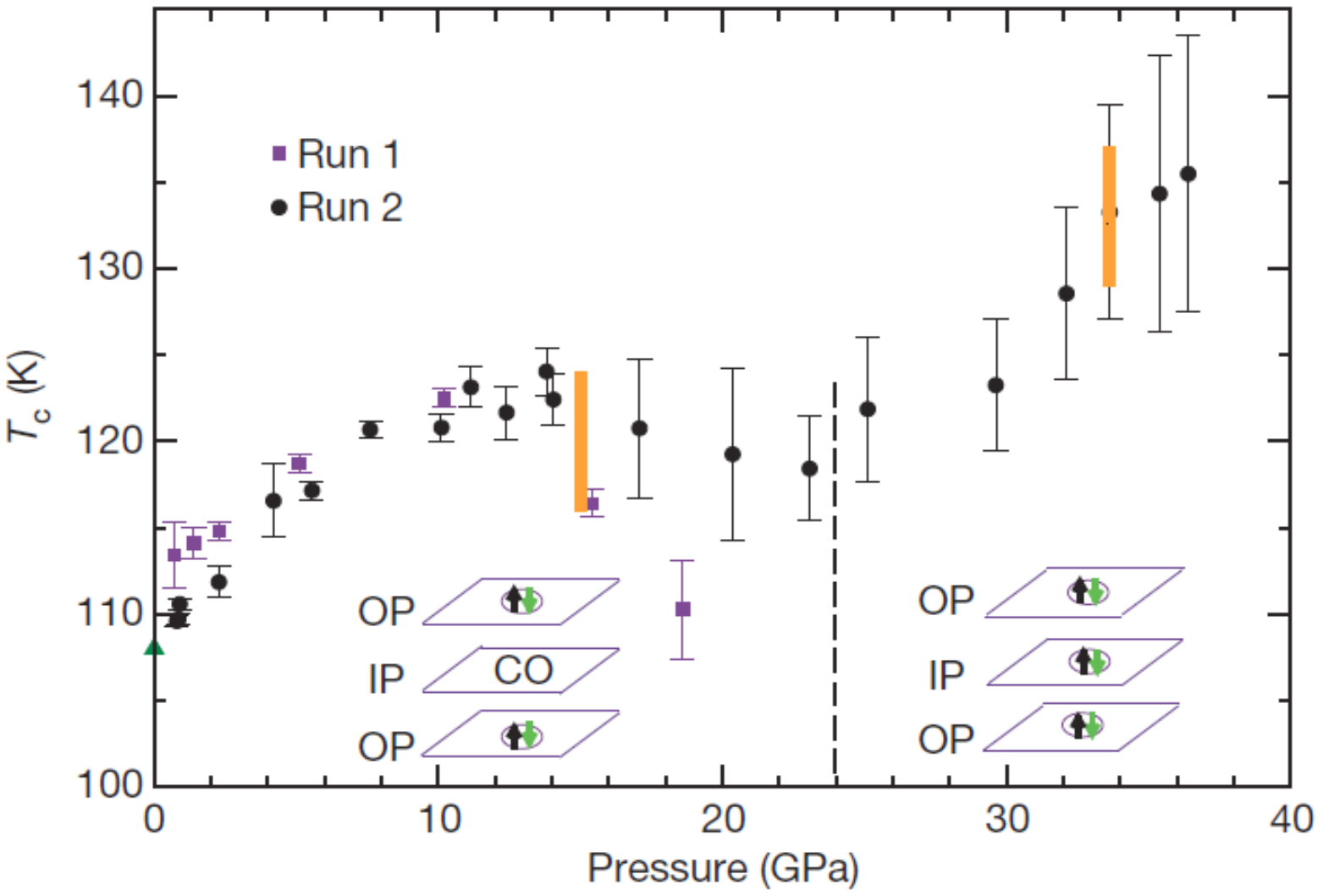

**Fig. 7** From X. J. Chen et al., Nature 466, 950 (2016), Fig. 3.

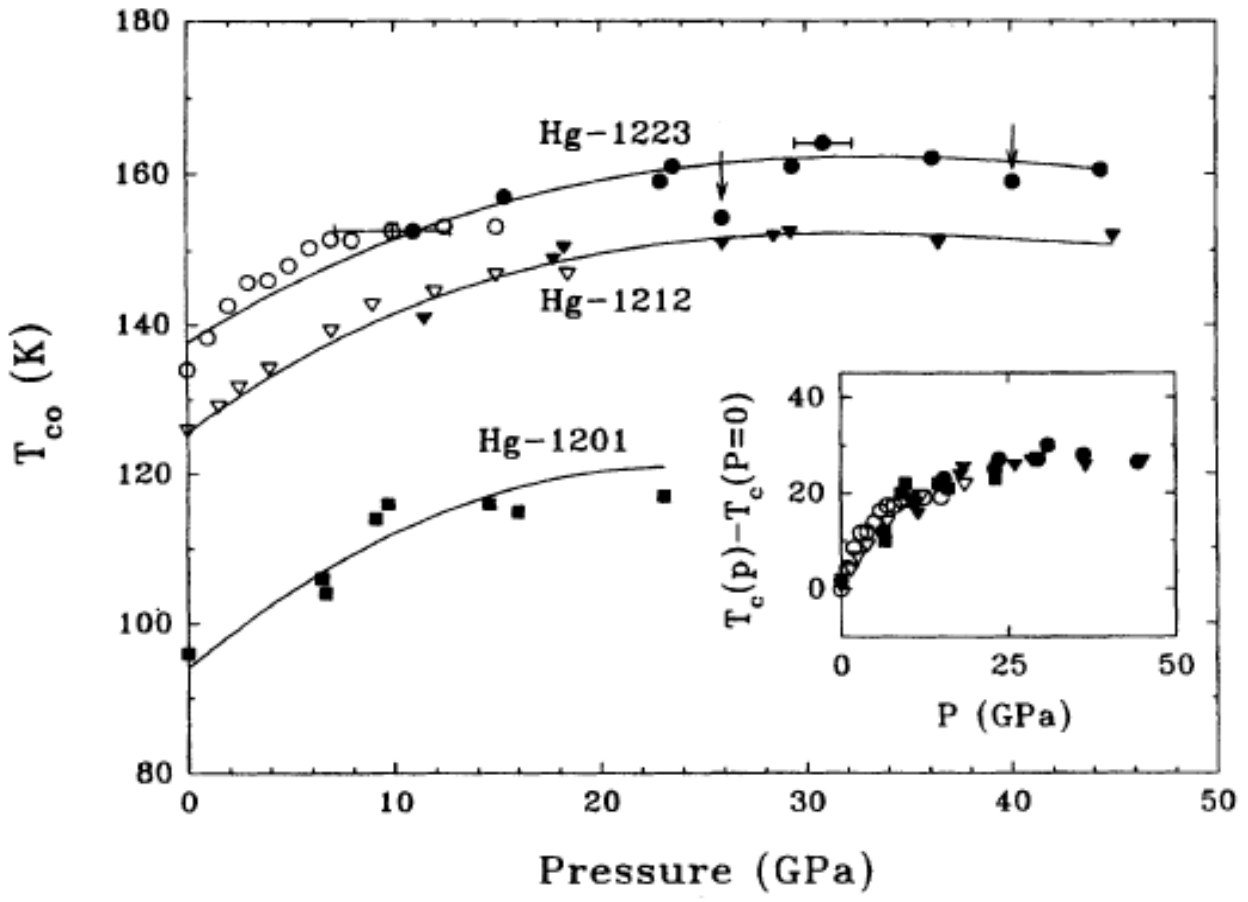

**Fig. 8** From L. Gao et al., Phys. Rev. B 50, 4260-4263 (R) (1994), Fig. 3.

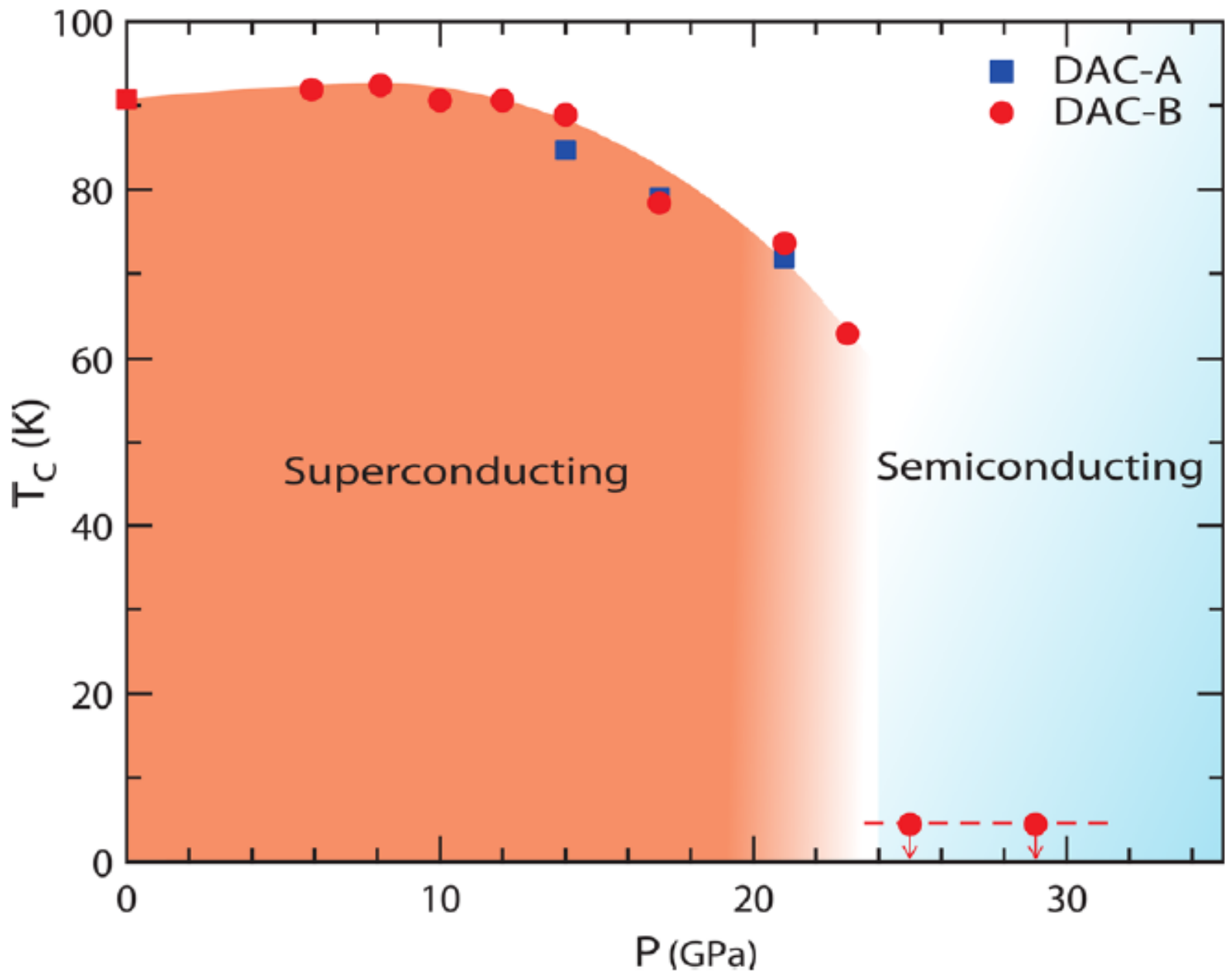

**Fig. 9** From T. Muramatsu, D. Pham, and C. W. Chu, Appl. Phys. Lett. 99, 052508 (2011), Fig. 3.

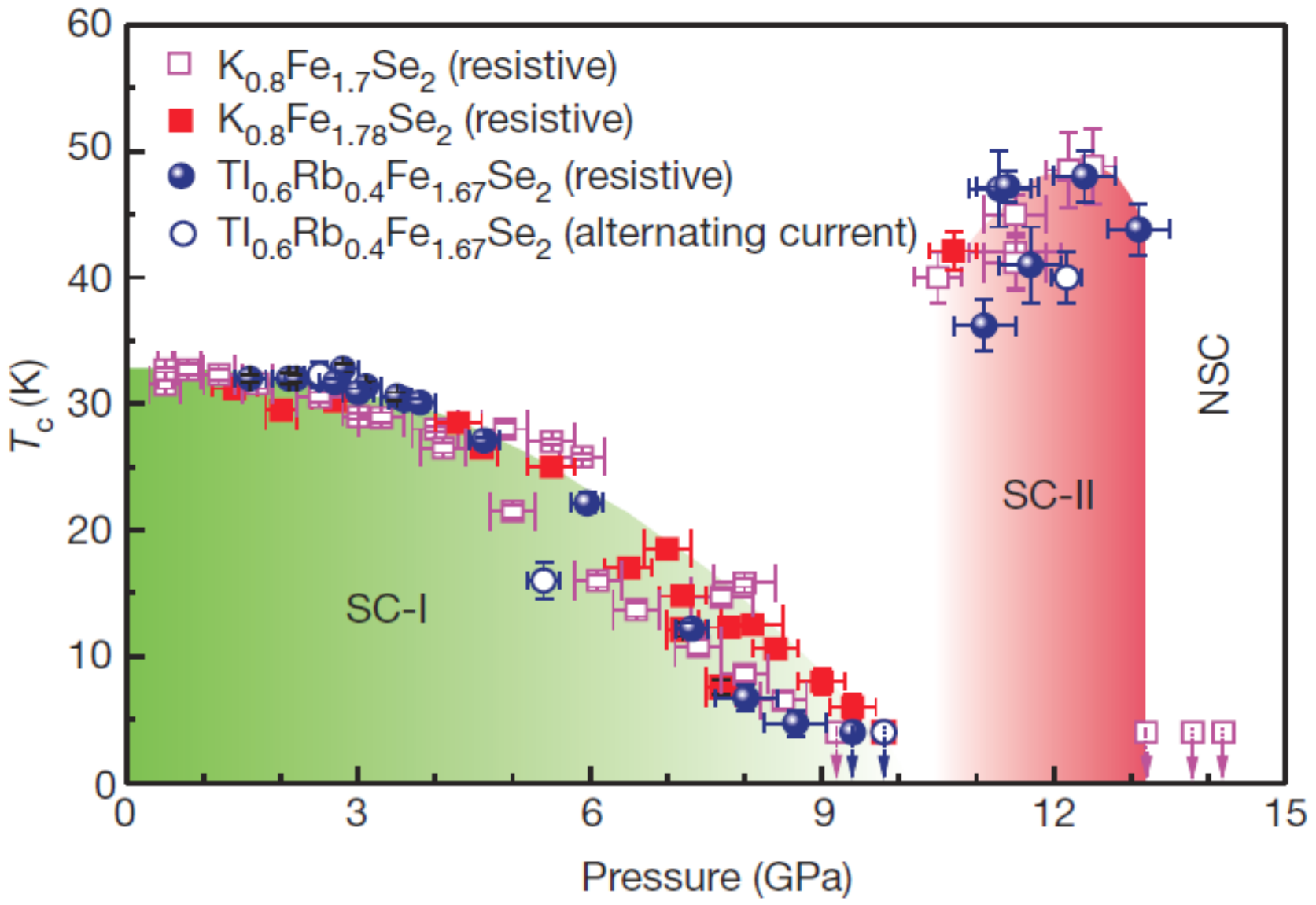

**Fig. 10** From L. L. Sun et al., Nature 483, 67 (2012), Fig. 4.

## 3. Hydrogen-Rich Molecular Compounds

The scientific significance of the recent report by Drozdov/Eremets et al. [18] is obvious and multifold: i. a new $T_c$ record of 203 K at ~ 200 GPa (vs. the current record of 164 K in HBCCO under 32 GPa set in 1993); ii. very high $T_c$ in a new class of 3D molecular solid materials (vs. the current 2D-cuprates for the last three decades); and iii. the possible viability of the so-called "first principle calculations" approach [19] to superconductors with higher $T_c$ through the electron-phonon interaction (vs. the enlightened empirical approach in the strong electron correlation systems [2]). In view of its significance, a careful and preliminary examination of the above claims has been made. We found that crucial questions remain and need to be clarified. In this respect, we are addressing two issues: 1. the existence of the Meissner state in $H_nS$; and 2. the viability of the current numerical calculations or "first principle calculations" of $T_c$ using different algorithms and codes. We plan to search for and study the Meissner effect in $H_3S$ under ~ 200 GPa below its $T_c$ ~ 200 K, as well as the high $T_c$ superconductivity predicted in Y-hydrides under pressures below 120 GPa: $YH_3$ ($T_c$ ~ 40 K at 17.7 GPa); $YH_4$ ($T_c$ ~ 84-95 K at 120 GPa); and $YH_6$ ($T_c$ ~ 252-264 K at 100 GPa) [20].

Although there are several excellent ultrahigh pressure labs in the US, China, Japan, and Europe, to date and to the best of our knowledge, there is only one lab (Shimizu's in Japan) that has reproduced the resistive transition in $H_3S$ together with Eremets [21], and none has reported the magnetic transition claimed by Eremets et al. [18], although a ZFC transition by nuclear resonating scattering was later reported by Troyan/Eremets et al. [22] and by an ac susceptibility technique by Huang/Cui [23]. Indeed, the large resistance drop, the negative magnetic field effect, and the isotope effect are consistent with a superconducting transition. However, given the extreme challenges in the resistive and magnetic experiments under ultrahigh pressure, and the fact that similar field and isotope effects can also be observed in some metal-semiconductor transitions of, e.g., transition-metal oxides, a scenario has also been proposed in which the sample undergoes a non-superconducting Keldysh-Kopaev transition at high temperature and is superconducting only below 40 K, after analyzing the resistive data and comparing them with those of HTS cuprates, pnictides, and organic superconductors [24]. We have therefore analyzed in detail all magnetic data published for the Meissner effect, whose existence below 203 K will be a sufficient proof of superconductivity in $H_3S$ under ultrahigh pressures, and have concluded that its existence remains an open question. We therefore are in collaboration with Eremets to search for it by deploying a sensitive magnetic technique as shown in Fig. 11 for the search. We will also examine the $YH_n$, which have been predicted to be superconducting below 150 GPa, to test the predictability of the so-called "first principle calculations".

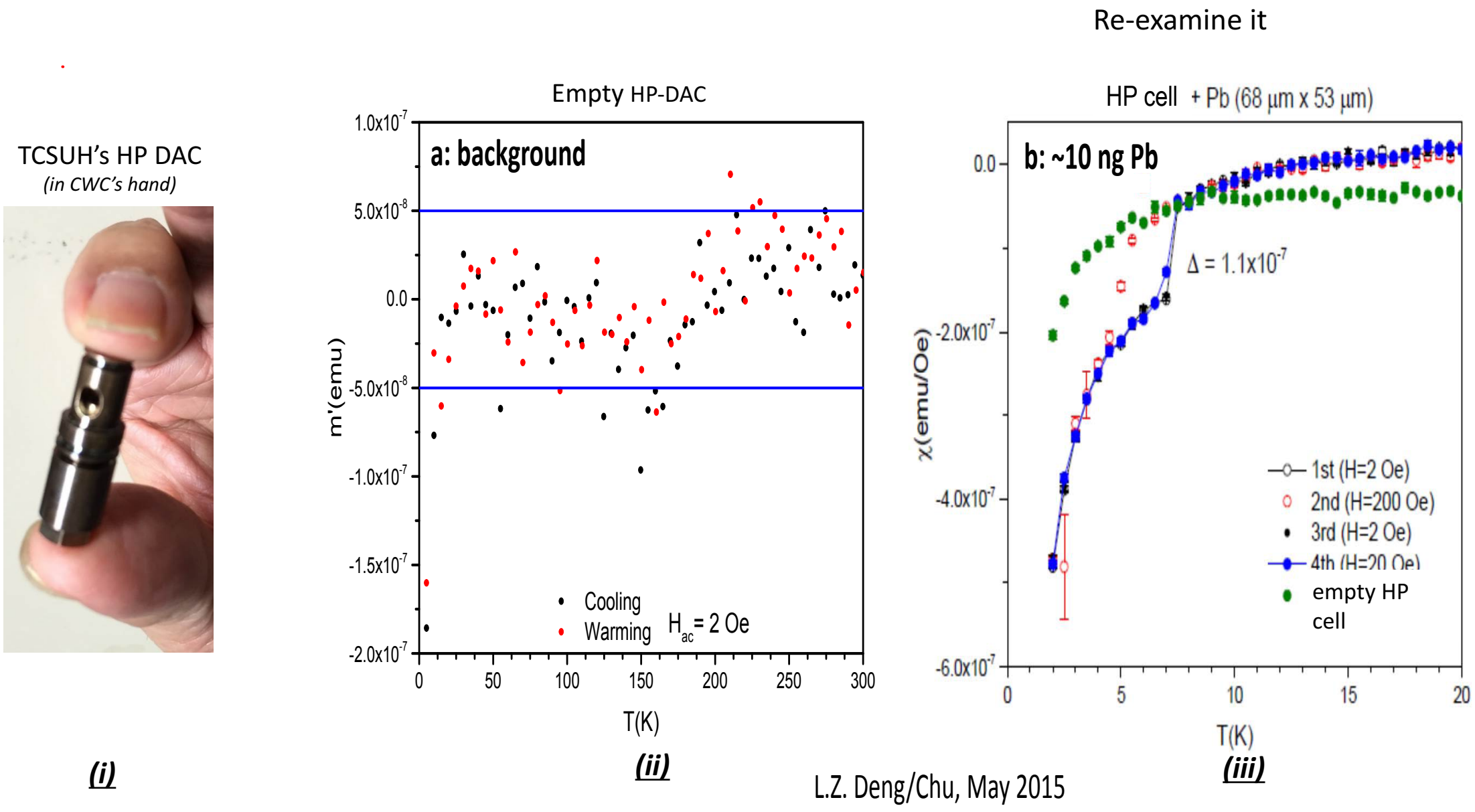

**Fig. 11** From Deng/Eremets and Chu.

## 4. Optimization of Multi-interaction in Multiferroics

As discussed in Ref. 2, known high $T_c$ superconductors exhibit multiple interactions, and magnetic interaction has been considered by many to play an important role. The question arises whether ferroelectric interaction can play a beneficial role for superconductivity as depicted in Fig. 12. Given the high ferroelectric or antiferroelectric ordering temperature, a beneficial effect on superconductivity similar to that of magnetism may not be impossible.
Multiferroics exhibit multiple competing interactions and display concurrently various ground states, e.g. the coexistence of magnetism and ferroelectricity. Studies by us and others show that ferroelectricity in multiferroics can be induced by magnetism, pressure, or an external magnetic field. It has also been shown recently that superconductivity can coexist with ferromagnetism and that magnetic field can induce superconductivity. This raises the question whether superconductivity can evolve from ferroelectricity directly or indirectly through magnetism. Since multiferroics belong to the class of highly correlated electron material systems that possess transition temperatures up to above room temperature, it is conjecture that it may not be impossible to achieve superconductivity with a very high $T_c$ by optimizing the various interactions present. The first order of business is to metallize the multiferroics. The chemical doping approach fails because of the high stability of the chemical bonding. We therefore have decided that the most effective way is through high pressure. Unfortunately, no detection of superconductivity has been achieved by us, to date. We have attributed the failure possibly to the Mn element that often appears in the multiferroics, since Mn is a killer of superconductivity (to the best of my knowledge there exist only two low $T_c$ superconducting Mn compounds, including one recently discovered by Cava). We are now looking at multiferroics without Mn. Preliminary results appear encouraging, as demonstrated in Fig. 13.

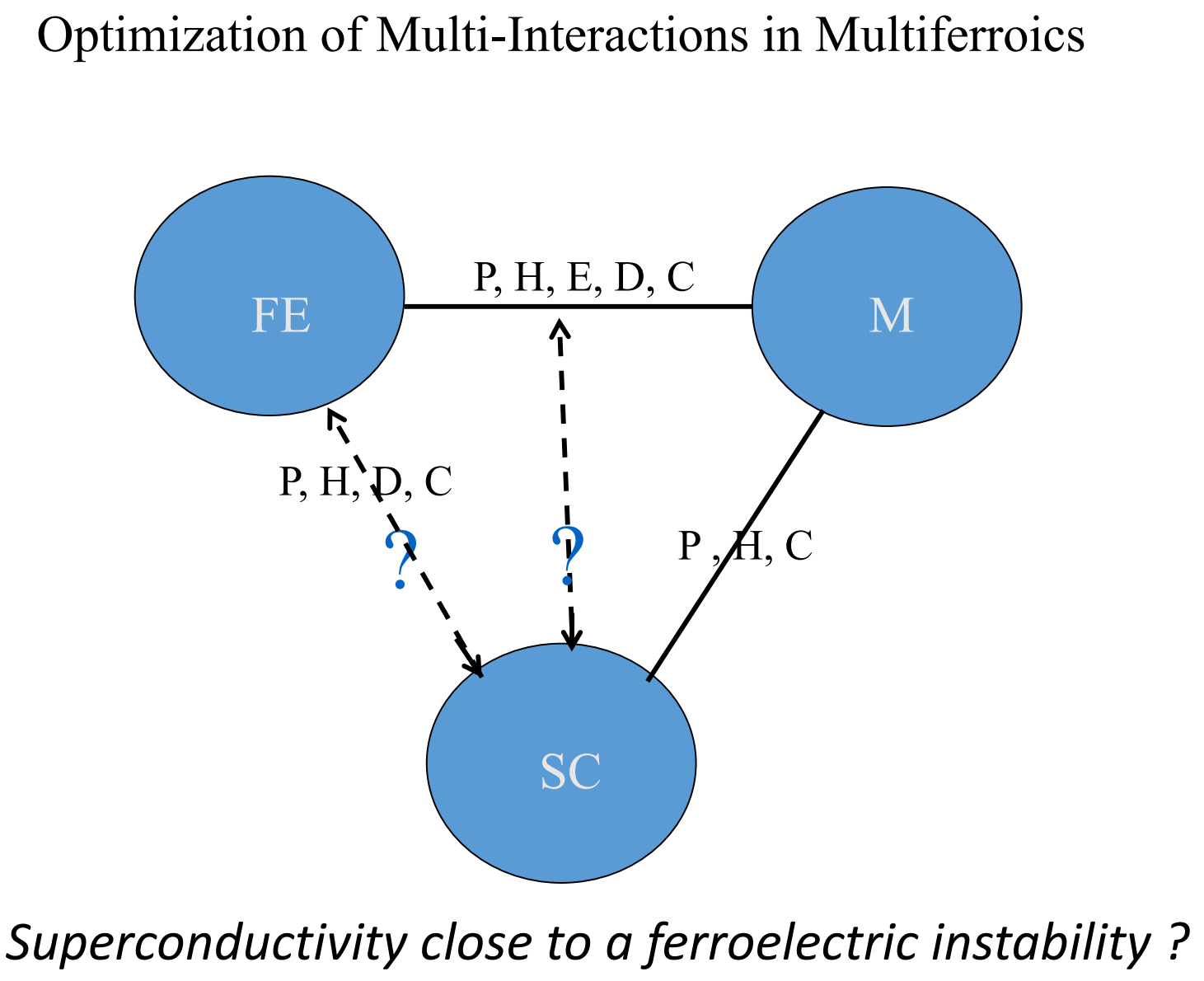

**Fig. 12** From C. W. Chu, AAPPS Bulletin 18, 9 (2008), Fig. 3.

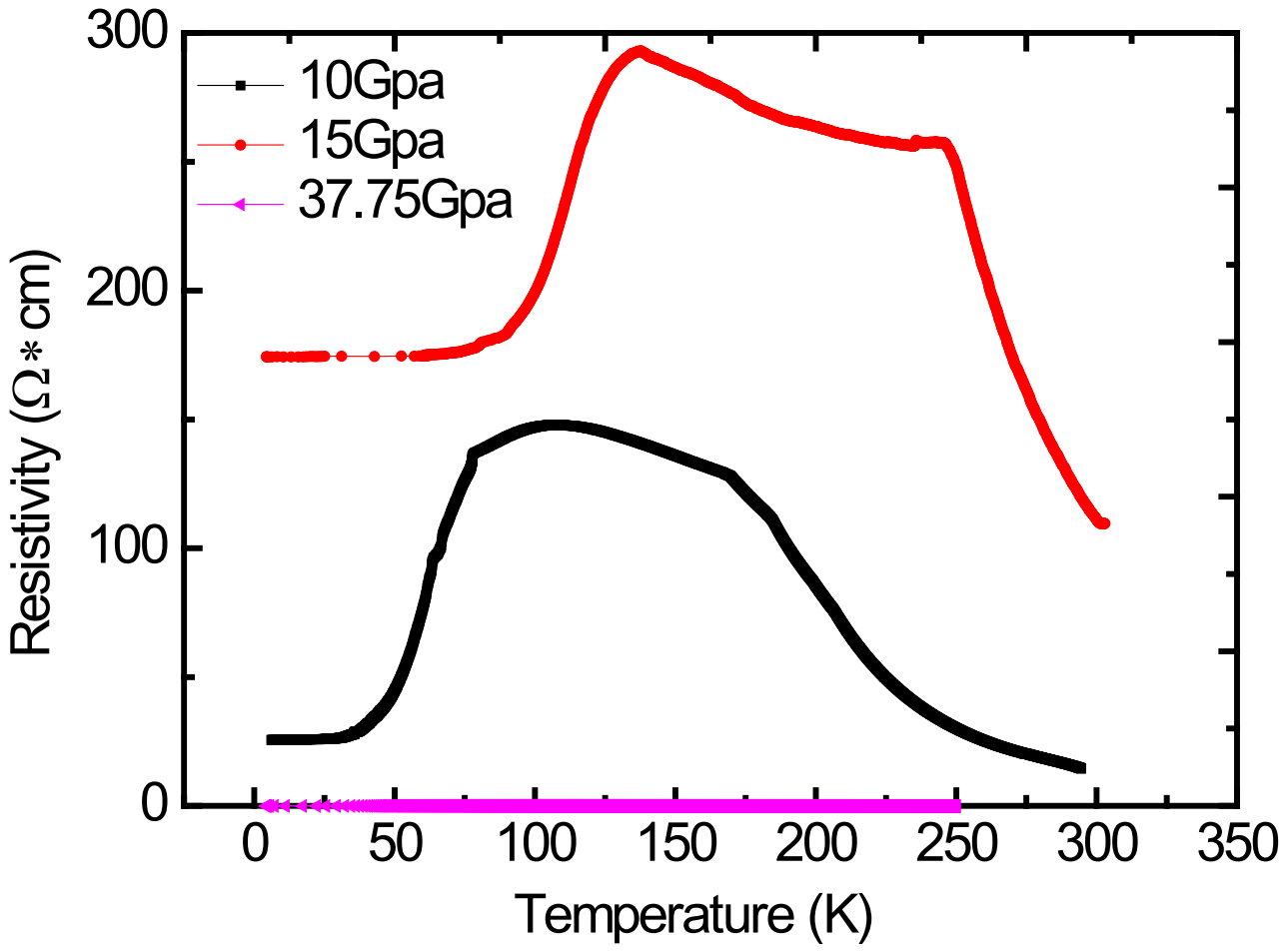

**Fig. 13** A multiferroic without Mn. From Z. Wu, S. Y. Huyan, L. Z. Deng/C. W. Chu.

As is evident, this presentation represents only a progress report of the effort to search for higher $T_c$ with great hope at Houston. I understand that vision does not count until real exciting results emerge. We will keep the faith in all of our efforts. With some luck, together we shall reach the 300 K wonderland as an ultimate tribute to Alex.

**Acknowledgments**

I would like to express my gratitude to current and former students and colleagues at TCSUH, B. Lv, Z. Wu, Y.Y. Xue, L.Z. Deng, S.Y. Huyan, H.M. Yuan, B. Lorenz, M. Gooch, F.Y. Wei, Y.Y. Sun, J.K. Meen, K. Zhao, L. Gao, F. Chen, T. Muramatsu, and D. Pham, for technical assistance and discussion. The work in Houston is supported in part by US Air Force Office of Scientific Research Grant FA9550-15-1-0236, the T. L. L. Temple Foundation, the John J. and Rebecca Moores Endowment, and the State of Texas through the Texas Center for Superconductivity at the University of Houston.

**References**

1. J. G. Bednorz and K. A. Müller, Z. Phys. B 64, 189 (1986).
2. C. W. Chu, AAPPS Bulletin 18, 9 (2008).
3. R. Kleiner, F. Steinmeyer, G. Kunkel, and P. Müller, Phys. Rev. Lett. 68, 2394 (1992).
4. D. Allender, J. Bray, and J. Bardeen, Phys. Rev. B 7, 1020 (1973).
5. J. Pereiro, A. Petrovic, C. Panagopoulos, and I. Bozovic, Phys. Express 1, 208 (2011).
6. B. Lv et al., Proc. Natl. Acad. Sci. USA 108, 15705 (2011); S. R. Saha et al., Phys. Rev. B 85, 024525 (2012).
7. L. Z. Deng et al., Phys. Rev. B 93, 054513 (2016).
8. H. Xiao et al., Phys. Rev. B 85, 024530 (2012).
9. S. Ran et al., Phys. Rev. B 83, 144517 (2011); B. Saparov et al., Sci. Rep. 4, 4120 (2014).
10. K. Zhao et al., Proc. Natl. Acad. Sci. USA 113, 12968 (2016).
11. A. Kreyssig et al., Phys. Rev. B 78, 184517 (2008).
12. L. Gao et al., Phys. Rev. B 50, 4260(R) (1994), and references therein.
13. X. J. Chen et al., Nature 466, 950 (2010), and references therein.
14. T. Muramatsu, D. Pham, and C. W. Chu, Appl. Phys. Lett. 99, 052508 (2011).
15. M. R. Presland et al., Physica C 176, 95 (1991).

16. D. H. Lee, Chin. Phys. B 24, 117405 (2015).
17. L. L. Sun et al., Nature 483, 67 (2012).
18. A P. Drozdov et al., Nature 525, 73 (2015).
19. See for example, “Materials by Design: An NSF-Sponsored Workshop at UC Santa Barbara”, March 17-19, 2011, http://www.mbd.mrl.ucsb.edu.
20. D. Y. Kim, R. H. Scheicher, and R. Ahuja, Phys. Rev. Lett. 103, 077002 (2009); Y. W. Li et al., Sci. Rep. 5, 9948 (2015).
21. M. Einaga et al., Nature Physics 12, 835 (2016).
22. I. Troyan et al., Science 351, 1303 (2016).
23. X. L. Huang et al., arXiv:1610.02630v1 [cond-mat.supr-con] (2016).
24. L. S. Mazov, arXiv:1510.00123v1 [cond-mat.supr-con] (2015).